\documentclass[titlepage,twoside,12pt]{article}
\usepackage{amssymb}
\usepackage{amsfonts}
\begin{document}

{\LARGE \bf Comments on S Kauffman's \\ \\ paper arxiv:0907.2492} \\ \\

{\bf Elem\'{e}r E ~Rosinger} \\ \\
{\small \it Department of Mathematics \\ and Applied Mathematics} \\
{\small \it University of Pretoria} \\
{\small \it Pretoria} \\
{\small \it 0002 South Africa} \\
{\small \it eerosinger@hotmail.com} \\ \\

{\bf Abstract} \\ 

Deficiencies in Kauffman's proposal regarding a new way for building scientific theories are pointed out. A suggestion to overcome them, and in fact, independently construct mathematical theories which are beyond the reach of Goedel's incompleteness theorem is presented. This suggestion is based on bringing together recent developments in literature regarding inconsistent mathematics and self-referential mathematics. \\ \\

{\bf 1. Overshooting the Target} \\

Reductionist science is meant by Kauffman, [1], to be any theory which is built exclusively upon entailments involving efficient causes in the traditional sense of Aristotle. And in view of Goedel's incompleteness theorem, as well as Hawking's recent paper [2], Kauffman sees the necessity to propose a way aimed at going beyond reductionist
science. \\

Such a proposal, needless to say, is welcome. And it would be so independently of Goedel's result. After all, as long as one is pursuing science in proper ways there should not be a priori limitations on it, even if they may happen to come for strongly entrenched historical traditions. \\

The issue, therefore, is which may indeed be certain proper ways for science beyond the one called reductionist in [1]. \\

In this regard, one can note that the analysis in [2] is rather naive, since it interprets Goedel's incompleteness as a consequence of the self-referential situation in which physicists are inevitably part of the physical realms. As for going beyond what in [1] is called reductionist science, the suggestion in [2] simply consists in a succession of reductionist theories of physics, succession in which larger and larger sets of axioms are assumed, thus managing to some rather limited extent to overcome Goedel's incompleteness. \\

From the point of view of such a rather trivial suggestion, the proposal in [1] is considerably more sophisticated and deep. Yet, as argued in the sequel, it is highly questionable on several grounds, and above all, in view of the fact that it cannot secure ways which could be seen as representing in any more proper manner a rigorously enough scientific approach. \\

The basic model in [1] which inspires the specific proposal aimed at going beyond reductionist science is taken from biology, and specifically, from Darwin's view of the evolution of species. The way Darwin's view is understood in [1] is that evolution in biology is not - and simply, cannot be - merely a process based on the action of efficient causes alone. Instead, it is the effect of an interplay between laws which act as enabling constraints, and on the other hand, efficient causes which become operational within the respective constraints, with both constraints and causes partly interacting and evolving, as if according to, what is called, some "blind final cause". \\

Let us now consider to what extent such an understanding of Darwinian evolution in biology may indeed be an appropriate method for building future scientific theories, instead of the present method in which mathematical theories are axiomatically founded and then developed according to rules of logic, while theories of physics tend to be based on such mathematical theories. \\

An obvious feature of the proposal in [1] is that it can be significantly {\it loose} in the way efficient causes may come into play within the enabling constraints of laws. Furthermore, this looseness may considerably increase during the assumed interplay between the enabling laws and the efficient causes acting within their constraints. \\
Therefore, one can only wonder to what extent such a considerably loose process run by some assumed "blind final cause" can possibly lead to scientific theories which perform, when applied in technologies, with any kind of more appropriate precision, a precision for instance which Kauffman himself may, no doubt, require in the construction of any airplane he may ever wish to consider flying with ... \\

In biology, the results of such loose processes are not supposed to - and certainly, do not - lead to any scientific theory, but to individual living creatures. And an essential difference between a scientific theory and an individual living creature is, among others, the following. Such a living creature is biologically easily disposable. That is, the species to which it belongs need not necessarily vanish with the demise of that individual. And even if in some highly unusual situation the existence of a whole species may depend on the survival of that individual, the death of that individual is not likely to lead to the end of biosphere as a whole ... \\
Thus the biological viability, that is, the {\it validity} of such an individual as a biological entity, has hardly any risk attached to it outside of the respective individual. This is precisely why Darwinian evolution, to the extent that it operates as assumed in [1], has so far not come anywhere near to put an end to the whole of biosphere as such ... \\

In total contradistinction with such a practically {\it no risk} biological situation, a scientific theory, when accepted, introduces a considerable risk, and it does so from theoretical point of view, since it may become a component of further scientific theories. And it does even more so, as soon as certain technologies are developed involving it. \\
A living creature which turns out to be biologically invalidated, that is, it ends up by being selected off the biosphere does nowhere the same kind of massive harm an unvalidated scientific theory can cause, be it in the theoretical or practical realms. \\
Therefore, the {\it validity} of a scientific theory in mathematics or physics, among others, is of uppermost immediate concern, a concern of both theoretic and practical nature. Consequently, the issue of that validity simply cannot be left to the mercy of some "blind final cause" ... \\

It follows that the proposal in [1] is clearly overshooting the target by allowing far too many allegedly scientific theories to be built, without any systematic manner of validation, except for some future action of a "blind final cause" ... \\

In case theories generated as proposed in [1] were not aimed for mathematics or physics, but rather for, say, psychology, sociology, economics, or the like, then, as a rather unfortunate human tradition may illustrate it, they may not seldom turn out to be acceptable for quite a while, that is, until such time when their nefarious consequences would become far too obvious for far too many ... \\

However, as far as mathematics, physics, chemistry, or for that matter, molecular biology are concerned, the validity of scientific theories should rather be built upon the well known criteria of rigor and precision in their construction, as well as present testing, instead of that left to some assumed "blind final cause" ... \\ 

A further major deficiency of the proposal in [1] is in its total lack of any kind of specific insight into the {\it micro} or {\it inner} aspects of the ways scientific theories end up being effectively constructed. Indeed, the proposal in [1] only sets up a {\it macro} type framework for the production of scientific theories, namely, through the mentioned interplay between laws which act as enabling constraints for certain efficient causes. And clearly, such a setup cannot give more insight into the specifics of the effective construction of scientific theories than an {\it outer} observer of the respective process could obtain. \\
Indeed, we are required by the proposal in [1] simply to {\it abandon} completely the present way scientific axiomatic theories are constructed step by step, that is, through a process which in its rigorous formal aspect offers a considerable transparency. \\

In conclusion, the proposal in [1] as it stands is considerably unrealistic when it aims to deliver scientific theories. \\
However, it is at the same time important to realize that a suitable {\it completion} of the proposal in [1] can possibly have significant merit. And such a completion may involve the following two additions : \\

1) The extreme abundance - referred to above as overshooting - in the production of alleged scientific theories, as inevitably follows from the present form of the proposal in [1], should be suitably curtailed. \\

2) The micro, or inner specific aspects in the construction of scientific theories should be specified in a clear, rigorous and transparent way. \\ \\

Related to the above second point, we make now certain suggestions. \\ \\  

{\bf 2. Nearer to the Target} \\

As mentioned, the concern with going beyond the present day axiomatic theories in mathematics or physics has lately gained some attention. \\
The suggestion in [2] seems to be too simple, and certainly, it does not address in any way the deeper reasons why usual axiomatic theories in mathematics or physics fall prey to Goedel's incompleteness theorem. \\
On the other hand, the proposal in [1] does open up a far larger possible pool for scientific theories, and in fact, as argued above, does so excessively. Furthermore, that proposal, similar to [1], fails to consider and identify those essential features of usual axiomatic theories in mathematics or physics which make Goedel's incompleteness theorem apply to them. \\

Here, following [3-6], we suggest a way to implement the requirement 2) at the end of the previous section. It should be noted that, as seen next, the implementation suggested here offers the construction of axiomatic mathematical theories which go far beyond the usual, presently used ones. Consequently, it can all alone, and by itself give a significant departure beyond the mathematical theories which happen to be subjected to Goedel's incompleteness theorem. \\

Also, this suggestion can be used in conjunction with the proposal in [1], as soon as that proposal is amended according to the requirement in 1) at the end of the previous section. \\

And now, to our suggestion based on [3-6], which is further based on [7-9]. \\

Historically, there has since times immemorial been an absolute rejection of systems of thought which lead to logical contradiction. \\
One of the effects of that total rejection has been the rejection of self-referential logical constructs, since as know from ancient times, they often led to contradictions, as illustrated in the famous ancient Greek paradox of the liar. \\

Recently however, [7,4], there has been an interest in inconsistent mathematics, that is, mathematics built upon a contradictory set of axioms. \\

Amusingly in this regard, in spite of the mentioned absolute rejection of logical contradictions, and especially in mathematics, we have nevertheless kept more and more indulging ourselves precisely in such mathematics ever since the use of digital electronic computers. \\
Indeed, albeit missed by nearly everyone, such computers are inevitably functioning based on the following system of axioms, when considered operating upon integers : \\

1) The Peano Axioms \\

plus \\

2) The Machine Infinity Axiom \\

which is usually formulated as follows : \\

(MIA)~~~ $ \exists ~ M >> 1 ~:~ M + 1 = M $ \\

where typically, one may have $M = 10^{100}$. \\

And clearly, the system of axioms 1) plus 2) is contradictory. \\

Yet we do not hesitate to use our digital electronic computers more and more massively, and even fly in airplanes designed and constructed with their essential help. \\

Details, albeit rather elementary, regarding the safe use of such contradictory axiomatic systems can be found in [7,4]. \\

So much for the traditional "no-go" regarding logical contradictions. \\

As for the other traditional "no-go", namely, of self-referential logical constructs, it happened in the early 1980s that the use of digital electronic computers led to yet another fundamental reconsideration, [8,9]. Namely, far reaching research was undertaken in foundations of a set theory along lines which contain a significant amount of self-referential definitions, yet, unlike the usual set theory with the Russell paradox, do not lead to any contradictions, provided that usual set theory does not do so either. \\

As it happens so far, this self-referential mathematics has on purpose been developed only with the aim to be consistent, provided that the usual set theory is so, [8,9,6]. And as can be seen, even within the confines of that restriction, it leads to a most remarkable extension of usual set theory, and thus of mathematics as such. \\

As a further step, therefore, one may consider the joining of inconsistent mathematics with self-referential mathematics. And in doing so, it may well happen that the resulting theories would no longer be within the realms of      Goedel's incompleteness theorem. \\
 
As for the possibility of that happening, we can recall that the proof of Goedel's incompleteness theorem does essentially involve a self-referential argument, brought about by the arithmetisation through the celebrated Goedel numbering of the propositions of Peano arithmetics. Thus to the extent that one goes over to mathematical theories which from the start are self-referential, the situation with the Goedel argument may change. Furthermore, the operative effect of the Goedel argument is that it leads to the dichotomy that either Peano arithmetics is inconsistent, or it is incomplete. Hence, once one is no longer rejecting inconsistent mathematical theories, the mentioned conclusion of incompleteness of Goedel's respective argument may fall away.

\end{document}